\documentclass[10pt]{article} 
\usepackage{amsmath,amsfonts}
\usepackage{algorithmic}
\usepackage{algorithm}
\usepackage{array}
\usepackage[caption=false,font=normalsize,labelfont=sf,textfont=sf]{subfig}
\usepackage{textcomp}
\usepackage{stfloats}
\usepackage{url}
\usepackage{verbatim}
\usepackage{graphicx}
\usepackage{cite}
\usepackage[colorlinks=true,linkcolor=black,anchorcolor=black,citecolor=black,filecolor=black,menucolor=black,runcolor=black,urlcolor=black]{hyperref}

\usepackage{booktabs}
\usepackage{multirow}

\usepackage{xcolor}

\usepackage[margin=1in]{geometry}
\vspace{-30 cm}

\begin{document}


\title{A stochastic noise model based excess noise factor expressions for staircase avalanche photodiodes}

\author{Ankitha E. Bangera \\
\\
\it{Department of Electrical Engineering,} \\
\it{Indian Institute of Technology Bombay,} \\
\it{Mumbai$-$400076, India} \\
E-mail: ankitha\_bangera@iitb.ac.in; ankitha.bangera@iitb.ac.in}
\date{}

\maketitle
\thispagestyle{empty}

\begin{abstract}
Multistep staircase avalanche photodiodes (APDs) are the solid-state analogue of photomultiplier tubes, owing to their deterministic amplification with twofold stepwise gain via impact ionization. Yet, the stepwise impact ionization irregularities worsen with increasing step counts, which are a major source of internal noise in these APDs. Some noise models for staircase APDs have been previously reported, where the excess noise factor expressions are based on Friis' noise factor formula for cascade networks, erroneously considering the power gains as the gains. Excess noise factor being a key component in staircase APDs' noise models, we formulate generalized excess noise factor expressions for multilayer graded-bandgap APDs in terms of their layer-wise ionization probabilities, applicable for all operating biases, which include the sub-threshold, staircase, and tunnelling breakdown regimes. We further derive simplified expressions for staircase APDs and prove that these expressions match Bangera's corrections to Friis' noise factor formulas for cascade networks. 
\end{abstract}

\section{Introduction}
\label{sec1}

In the past few decades, multilayer graded-bandgap avalanche photodiodes (APDs) operated in their staircase regime have become the semiconductor (or solid-state) analogue of photomultiplier tubes (PMTs) with deterministic amplification \cite{bib1,bib2,bib3,bib4,bib5,bib6}. Compared to PMTs, these solid-state devices have the advantages of being micro-sized, inexpensive, and operated at low voltages ($<$ 100V) \cite{bib1,bib2,bib3}. The conduction band profile in the energy-position band-diagram of these APDs appears similar to a series of steps when the applied bias ranges within the staircase regime \cite{bib1,bib2,bib3,bib5,bib6,bib7}. The function of these steps is similar to metallic dynodes in a PMT. Thus, these APDs operated in the staircase regime are referred to as multistep or $n$-step staircase APDs (or staircase APDs). However, the stepwise impact ionization irregularities worsen as the number of steps increases, giving rise to internal noises in these APDs \cite{bib2,bib5,bib10,bib11}. These solid-state devices could replace conventional PMTs if they are carefully designed to reduce these internal noises. 

Some noise models \cite{bib5,bib6,bib8,bib9,bib10} for staircase APDs have previously been reported. However, the total excess noise factor expression for $n$-stage staircase multipliers was first reported by Capasso and co-workers in the early 1980's \cite{bib5,bib6}. Capasso's excess noise factor expression and its equivalent forms are often used to analyze and fit experimentally measured data \cite{bib2,bib8,bib11}. Campbell and co-workers \cite{bib2,bib11} have recently reported the theoretical estimates of total excess noise factors using this formula. However, Capasso's formula and its related forms are based on Friis' noise factor (or noise figure) formulas for cascade networks \cite{bib8,bib9,bib12,bib13}, erroneously considering the power gains as the gains \cite{bib14}. Moreover, our previous work reports that Friis' noise factor formulas need a correction \cite{bib15}. Therefore, it is necessary to derive the correct excess noise factor expressions for staircase APDs. 

This article formulates generalized expressions for excess noise factors for multilayer graded-bandgap APDs in terms of their layer-wise ionization probabilities from the basic definition of excess noise factor for detectors. Our generalized expressions are applicable for all operating biases in a multilayer graded-bandgap APD, which include the sub-threshold, staircase, and tunnelling breakdown regimes. We further derive the expressions for stepwise and total excess noise factors for staircase APDs as a function of their stepwise ionization probabilities using a statistical random process method. For validation, we compare our theoretically estimated values with the measured total excess noise factor data for 3-step staircase APDs reported by March \textit{et al.} (Corrections \& amendments) \cite{bib2} and Dadey \textit{et al.} \cite{bib11}. We then demonstrate that our newly derived excess noise factor expressions for staircase APDs agree with Bangera's corrections to Friis' noise factor formulas for cascade networks.

\section{Theory} 
\label{sec2}

\subsection{Generalized excess noise factor expressions for multilayer graded-bandgap APDs} 
\label{subsec2_1}

\begin{figure}[!t]
\centering
\includegraphics[width=3.6in]{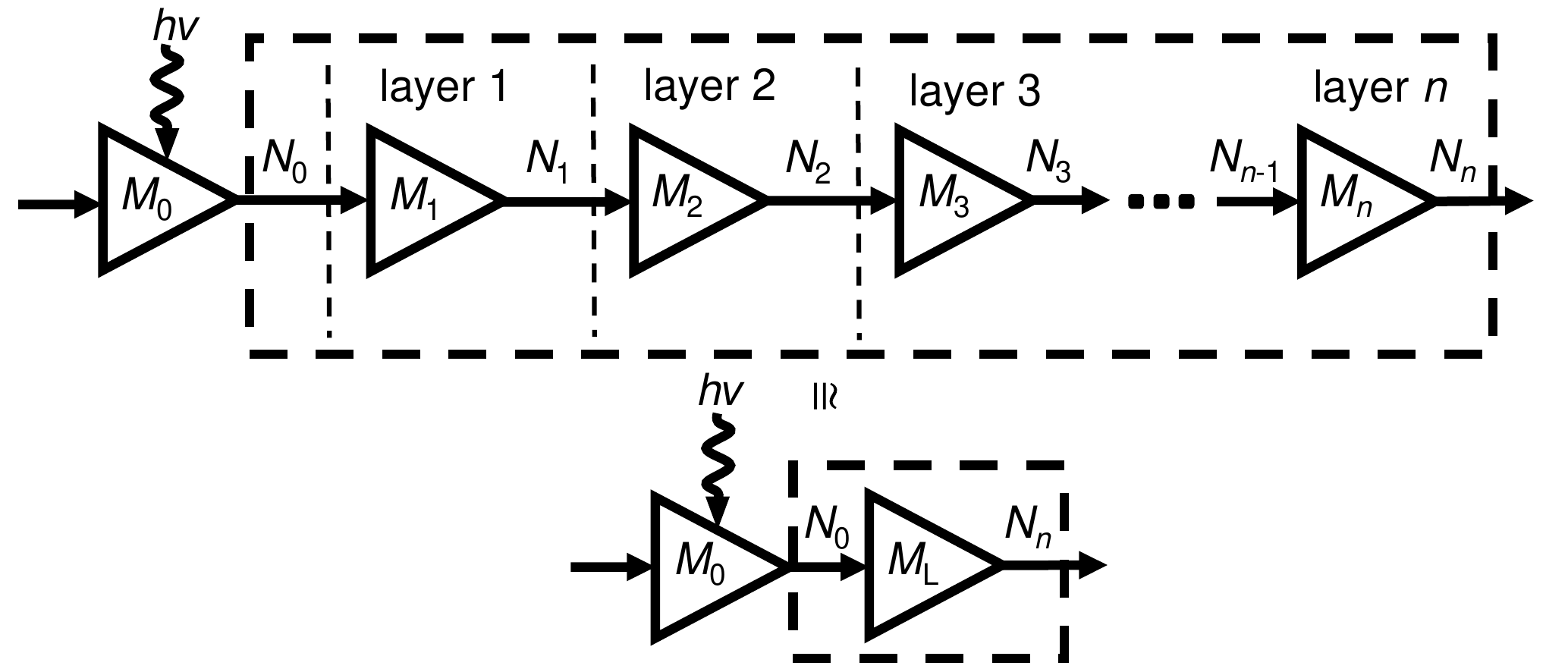}
\caption{Block diagram of an $n$-layer graded-bandgap APD forming a cascade amplifier. The photoelectron gain corresponding to the generation of photoelectrons in the absorption region is depicted as the first amplifier (left). The remaining amplifiers (top) correspond to layer-wise electron-multiplication gains at each graded-bandgap layer. The total layer-gain combines the effect of all these layer-wise gains represented as a single amplifier (bottom right).}
\label{fig_1}
\end{figure}

Owing to the layer-wise electron-multiplication observed in multilayer graded-bandgap APDs, these devices behave similar to cascade amplifiers. The block diagram in Fig.~\ref{fig_1} elaborates the amplification of photoelectrons in an $n$-layer graded-bandgap APD. Here, the first amplifier with gain $M_0$ represents photoelectron gain in the absorption region. $M_0=1$ indicates that each incident photon generates only one photoelectron and no electron-multiplication happens in the absorption region of APDs. However, the band-structure of multilayer graded-bandgap APDs appears discontinuous due to the heterojunction interfaces, even when the devices are unbiased \cite{bib1,bib5,bib6,bib7,bib10,bib11}. The average layer-wise gains represented as $M_x=\langle m_x \rangle$ correspond to free-electrons generated at the APD's $x$-th layer. From the block diagram in Fig.~\ref{fig_1}, total layer-gain ($M_\text{L}$) can be defined as the fraction of total electrons collected at the output of the last layer ($N_n$) per pulse input and total photoelectrons generated per pulse in the absorption region of the APD ($N_0$). Additionally, since these layers form a cascade structure, the mean of the total layer-gain of the staircase APD is equal to the product of all its layer-wise gains. Therefore,

\begin{equation}
\label{eqn_1}
M_\text{L} = \langle m_\text{L} \rangle =\prod_{x=1}^{n} \langle m_x \rangle
\end{equation}

The noise current spectral intensity (A$^2$Hz$^{-1}$) for these APDs is \cite{bib2,bib5,bib6,bib11},

\begin{equation}
\label{eqn_2}
\begin{aligned}
S^I (f) &=2q \langle m_{\text{T}}^2 \rangle I 
\end{aligned}
\end{equation}

where $q$ is the electron charge; $\langle m_{\text{T}} \rangle = M_{\text{T}} = M_0M_\text{L}$ is the overall average gain of the multilayer graded-bandgap APD; and $I$ is the current component, which includes the photo and dark currents. Due to extremely low electron-multiplication in the absorption regions of these APDs, the photoelectron gain $M_0\approx1$. Then, equation~\eqref{eqn_2} simplifies to, 

\begin{equation}
\label{eqn_3}
S^I (f)=2q \langle m_\text{L}^2 \rangle I
\end{equation}

The total excess noise factor of semiconductor devices (such as APDs) is defined as the ratio of the mean square value of its total gain to the square of its mean total gain \cite{bib2,bib4,bib9,bib10,bib11}, given by, 

\begin{equation}
\label{eqn_4}
F(M_{\text{T}}) = \frac{\langle m_\text{T}^2\rangle}{\langle m_\text{T}\rangle ^2} = 1+\frac{\text{var}{(m_\text{T}})}{\langle m_\text{T}\rangle ^2}
\end{equation}

In practical $n$-layer graded-bandgap APDs, an input electron at the $x$-th layer would generate one or more free-electrons due to impact ionization, or the input electrons might not impact-ionize \cite{bib1,bib2,bib5,bib6,bib8,bib9,bib10,bib11}. Furthermore, all the layers might not have equal ionization probabilities owing to the non-identically manufactured APD heterojunctions, the device operating bias, and so on \cite{bib2,bib11}. Since this event of layer-wise impact ionization is irregular (random), it can be considered a random process. Therefore, this article considers $X_x$ as a random variable for electron-multiplication at layer `$x$,' which represents the number of free-electrons generated by a single input electron at each layer with variable layer-wise ionization probabilities $p_{xi}$, where `$x$' indicates the layer number and `$i$' indicates the impact-ionized electron count per input electron, such that, 

\begin{equation}
\label{eqn_5}
\begin{aligned}
X_x \sim 
\begin{cases}
0 & 1-\left(\sum_{i=1}^{m}p_{xi}\right) \\
1 & p_{x1} \\ 
2 & p_{x2} \\
\vdots & \vdots \\
m & p_{xm} \\
\end{cases}
\end{aligned}
\end{equation}

Thus, $X_x$ indicated in equation~\eqref{eqn_5} would provide a generalized solution for all the layer-wise impact ionization irregularities observed in a practical $n$-layer graded-bandgap APD. Since the layer-wise ionization probabilities and the corresponding layer-wise gains are the characteristics of a layer, we consider the events of layer-wise impact ionization as statistically independent. Then, the mean total layer-gain of these APDs can be written as,

\begin{equation}
\label{eqn_6}
\begin{aligned}
M_{\text{L}} = \langle m_{\text{L}} \rangle = \bigg\langle \prod_{x=1}^{n} (1+X_x) \bigg\rangle = \prod_{x=1}^{n} \biggl( 1+ \sum_{i=1}^{m} \Bigl\{ (i)p_{xi} \Bigr\} \biggr) 
\end{aligned}
\end{equation} 

The total excess noise factor expression will be, 

\begin{equation}
\label{eqn_7}
\begin{aligned}
F(M_{\text{L}},p_{xi}) &= \prod_{x=1}^{n} \Biggl\{ \frac{\Bigl( 1+ \sum_{i=1}^{m} \bigl\{ [i(i+2)]p_{xi} \bigr\} \Bigr)}{\Bigl( 1+ \sum_{i=1}^{m} \bigl\{ (i)p_{xi} \bigr\} \Bigr)^2} \Biggr\} 
\end{aligned}
\end{equation} 

The layer-wise excess noise factor is expressed as, 

\begin{equation}
\label{eqn_8}
\begin{aligned}
F_x(M_{\text{L}},p_{xi}) = \frac{\Bigl( 1+ \sum_{i=1}^{m} \bigl\{ [i(i+2)]p_{xi} \bigr\} \Bigr)}{\Bigl( 1+ \sum_{i=1}^{m} \bigl\{ (i)p_{xi} \bigr\} \Bigr)^2}  
\end{aligned}
\end{equation} 

The detailed solutions for equations~\eqref{eqn_6}$-$\eqref{eqn_8} are in Appendix~\ref{secA2}. Equations~\eqref{eqn_6}$-$\eqref{eqn_8} are generalized expressions for a practical $n$-layer graded-bandgap APD, applicable for all operating biases. However, this article focuses on the staircase operation of multilayer graded-bandgap APDs, owing to their controlled layer-wise impact ionization with twofold layer-wise gain compared to the sub-threshold and tunnelling breakdown regimes.

\subsection{Staircase operation of multilayer graded-bandgap APDs} 
\label{subsec2_2}

\begin{figure}[!t]
\centering
\includegraphics[width=3.3in]{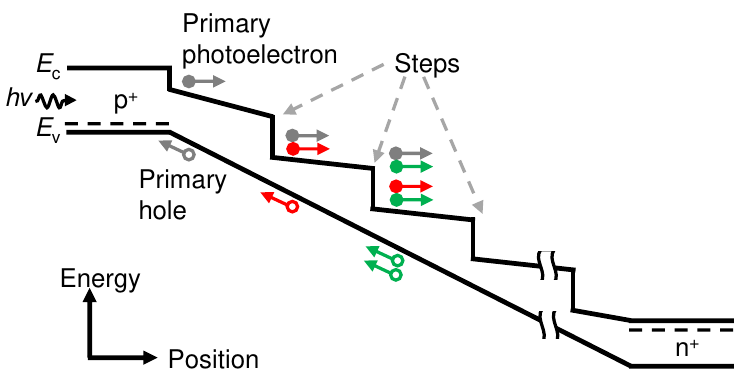}
\caption{Energy-position band-diagram of an ideal multistep staircase APD. The conduction band profile appears similar to a series of steps with a deterministic twofold stepwise gain at each step.}
\label{fig_2}
\end{figure}

Multilayer graded-bandgap APDs operated in their staircase regime are referred to as multistep or $n$-step staircase APDs (or staircase APDs). The conduction band profile in the energy-position band-diagram of staircase APDs appears discontinuous, similar to a series of steps \cite{bib1,bib2,bib3,bib5,bib6,bib7,bib10,bib11}. This article refers to these discontinuities as steps. Since the electron-multiplication is extremely low in the absorption regions of staircase APDs, the overall gain in these APDs will be $M_\text{T}\approx M_\text{S}$, where $M_\text{S}$ is the staircase gain corresponding to the total electron-multiplication at all steps. For an ideal $n$-step staircase APD, an input electron at each step will contribute to impact ionization by generating only one free-electron with unity ionization probability. Thus, the mean stepwise gain $\langle m_x \rangle=2$, where `$x$' is the step number. The mean staircase gain for these APDs will be a deterministic $2^n$ gain. Fig.~\ref{fig_2} depicts the energy-position band-diagram of an ideal multistep staircase APD.

However, in a practical $n$-step staircase APD, the stepwise impact ionization will remain a random process. But, the literature reports that the probability of input electrons generating more than one free-electron at each step is relatively low when these devices are operated in their staircase regime \cite{bib5,bib6,bib9,bib11}. Therefore, simplifying equation~\eqref{eqn_5} and neglecting the impact ionization events with lower probabilities, we consider that an input electron at step `$x$' would generate only one free-electron with stepwise ionization probability $p_x$. Then, $X_x$ is a random variable for electron-multiplication at step `$x$' with a Bernoulli distribution. Therefore,

\begin{equation}
\label{eqn_9}
\begin{aligned}
X_x \sim 
\begin{cases}
0 & (1-p_{x}) \\
1 & p_{x} \\ 
\end{cases}
\end{aligned}
\end{equation}

Again, the stepwise impact ionization events are statistically independent. Thus, the mean staircase gain expressed in terms of stepwise ionization probabilities $p_x$ is,

\begin{equation}
\label{eqn_10}
M_{\text{S}} = \langle m_{\text{S}} \rangle = \prod_{x=1}^{n}(1+p_{x}) 
\end{equation} 

The total excess noise factor expression will be, 

\begin{equation}
\label{eqn_11}
\begin{aligned}
F(M_{\text{S}},p_x) = \prod_{x=1}^{n} \biggl\{ \frac{(1+3p_x)}{(1+p_x)^2} \biggr\} = \prod_{x=1}^{n} F_x(M_{\text{S}},p_x) 
\end{aligned}
\end{equation} 

where $F_x(M_{\text{S}},p_x)$ is the stepwise excess noise factor, given by,  

\begin{equation}
\label{eqn_12}
\begin{aligned}
F_x(M_{\text{S}},p_x) = \frac{(1+3p_x)}{(1+p_x)^2} = \frac{\langle m_x^2 \rangle}{\langle m_x \rangle ^2}
\end{aligned}
\end{equation} 

Rearranging equation~\eqref{eqn_12}, we express the stepwise excess noise factors in terms of mean and variance of their stepwise gains, given by, 

\begin{equation}
\label{eqn_13}
\begin{aligned}
F_x(M_{\text{S}},p_x) &= 1 + \frac{p_x(1-p_x)}{(1+p_x)^2} = 1 + \frac{\text{var}(m_x)}{\langle m_x \rangle ^2}  
\end{aligned}
\end{equation}

The derivation and proof of equations~\eqref{eqn_10}$-$\eqref{eqn_13} are in Appendix~\ref{secA1}.

\section{Results and Discussion} 
\label{sec3}

Our newly derived expressions in terms of variable stepwise ionization probabilities $p_x$ are multidimensional plots. Therefore, to visualize these expressions, we consider equal stepwise ionization probabilities `$p$' at all `$n$' steps. Then, our mean staircase gain will be $\langle m_\text{S} \rangle=(1+p)^n$. From equation~\eqref{eqn_11}, the total excess noise factor expression will be,

\begin{equation}
\label{eqn_14}
F(M_{\text{S}},p) = \left[ \frac{(1+3p)}{(1+p)^{2}} \right]^n  
\end{equation}

\begin{figure*}[!t]
\centering
\includegraphics[width=4.5in]{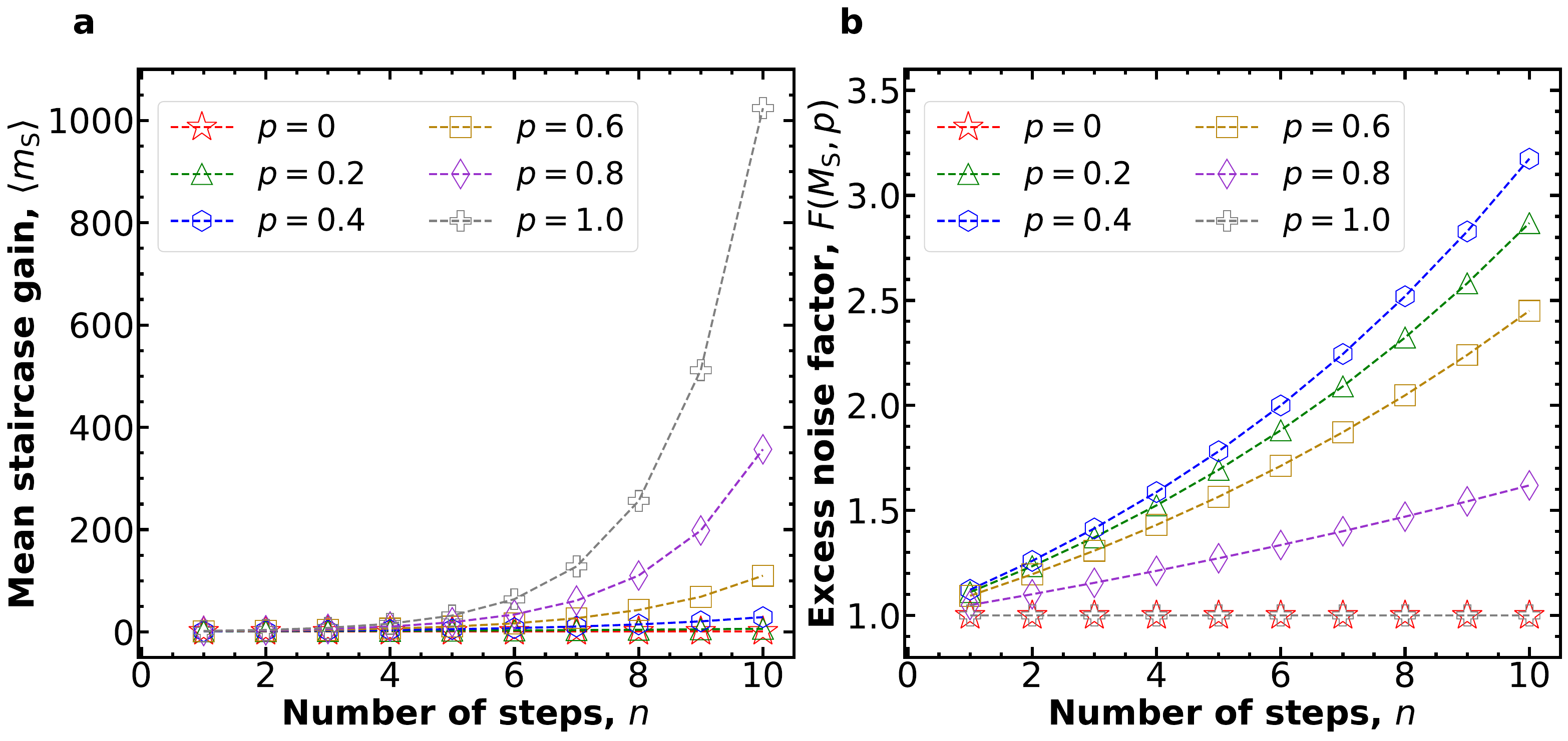}
\caption{Mean staircase gain and excess noise factor as a function of step count `$n$.' (a)$-$(b) The plot of staircase APD's mean staircase gain (a) and total excess noise factors (b) versus the step count `$n$' varying from 1 to 10, considering equal stepwise ionization probabilities `$p$.' Each curve corresponds to the value of `$p$' increasing from 0 to 1 with increments of 0.2.}
\label{fig_3}
\end{figure*}

Fig.~\ref{fig_3}a shows the mean staircase gain curves as an exponentially increasing function of step count `$n$.' The $p=1$ curve predicts an ideal $2^n$ behavior in staircase APDs with a deterministic twofold stepwise gain in the absence of internal noises. The $p=0$ curve is a horizontal line at $\langle m_\text{S} \rangle = 1$, which predicts a unity gain with no amplification at the steps. Fig.~\ref{fig_3}b shows a family of non-saturating curves corresponding to total excess noise factors, traced by equation~\eqref{eqn_14}. These curves vary as exponentially increasing functions of step count, except for $p=$ 1 and 0 curves that are horizontal lines at $F(M_\text{S},p)=1$, indicating the absence of internal noise with a deterministic $2^n$ staircase gain ($p=1$ condition) or nil stepwise impact ionization with unity staircase gain ($p=0$ condition). Here, we observe that the curves for total excess noise factor initially progress as the `$p$' is increased from 0 to 0.4, after which the curves appear damped. This is because the total excess noise factor peaks at ionization probability $p\approx0.3$.

\begin{figure*}[!t]
\centering
\includegraphics[width=4.5in]{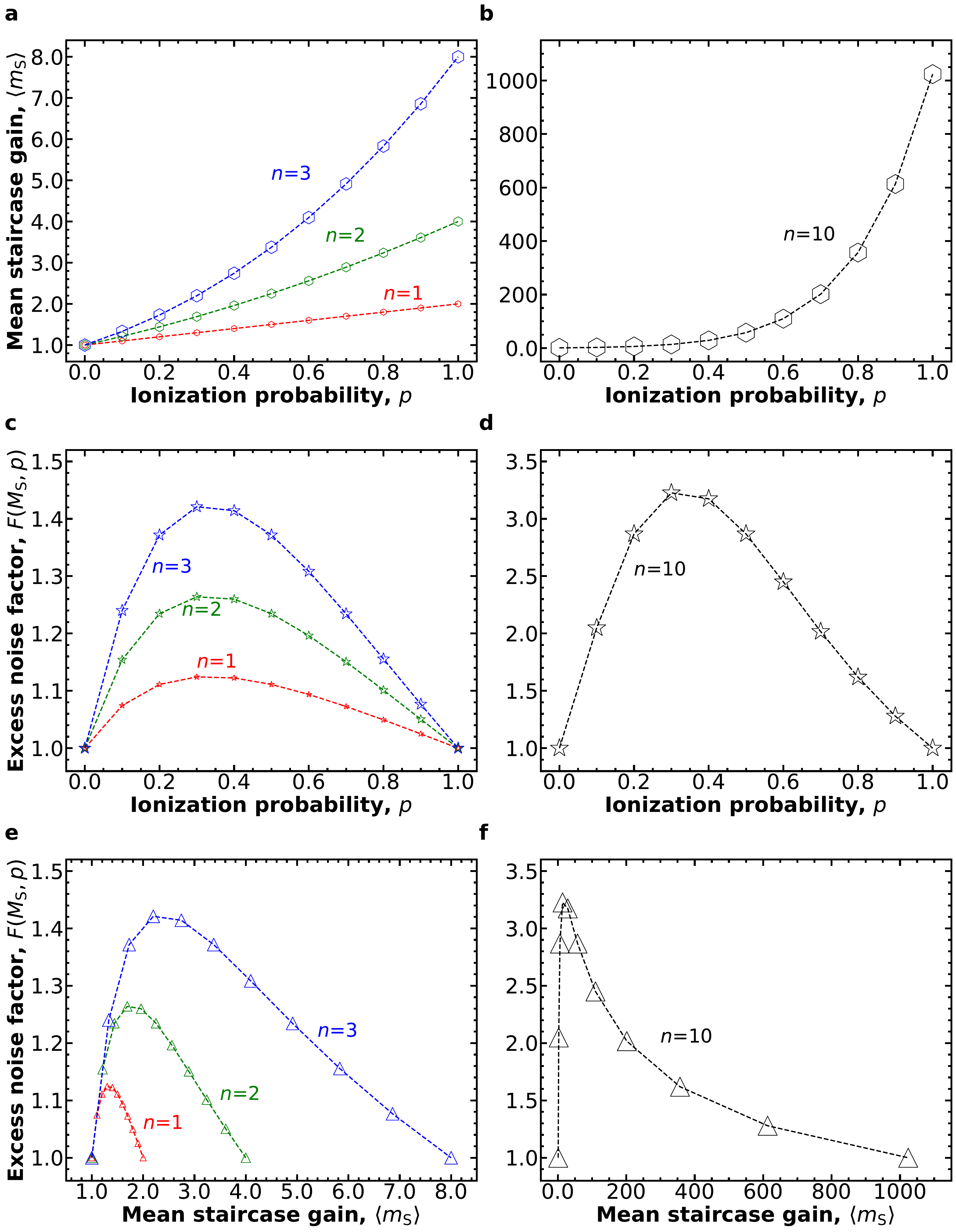}
\caption{Mean staircase gain and excess noise factor as a function of equal stepwise ionization probability `$p$.' (a)$-$(b) The plot of total excess noise factors for one-, two-, and three-step staircase APDs (a) and for a ten-step staircase APD (b) versus stepwise ionization probabilities `$p$'. (c)$-$(d) The plot of mean staircase gain for one-, two-, and three-step staircase APDs (c) and for a ten-step staircase APD (d) in terms of equal stepwise ionization probabilities `$p$'. (e)$-$(f) Curves of total excess noise factor of one-, two-, and three-step staircase APDs (e) and a ten-step staircase APD (f) as a function of mean staircase gain.}
\label{fig_4}
\end{figure*}

The mean staircase gain expressed as an $n$-th order power function of ($1+p$) is shown in Fig~\ref{fig_4}a and Fig~\ref{fig_4}b. The total excess noise factor as a function of `$p$' is shown in Fig~\ref{fig_4}c and Fig~\ref{fig_4}d. Here, all the curves peak at $p\approx0.3$, indicating that the average maximum internal noise due to irregular stepwise impact ionization in staircase APDs occurs when the stepwise ionization probability is approximately 0.3. Fig~\ref{fig_4}e and Fig~\ref{fig_4}f depict the total excess noise factor for staircase APDs as a function of mean staircase gain. Here, the curves' peaks right-shift with step count.

\subsection{Experimental validation} 
\label{subsec3_1}

The total excess noise factor value for a 3-step staircase APD with an average staircase gain of 7.24 extracted from the curves in Fig.~\ref{fig_4}e is approximately 1.05. Measurements by Dadey \textit{et al.} \cite{bib11} at an average staircase gain of 7.24 show that the measured total excess noise factor for 3-step staircase APDs fabricated by them ranges from approximately 1.1125 to 1.0375, with a mean value of 1.08. Our expression's result of 1.05 falls within the measured range given by Dadey \textit{et al.} and is closer to their measured average value of 1.08. Similarly, March \textit{et al.} (Corrections \& amendments) \cite{bib2} report a measured total excess noise factor of approximately 1.08 at an average staircase gain of 7.24 for 3-step staircase APDs fabricated by them, which agrees well with our expression's estimate of 1.05. Remarkably, these evidences validate our total excess noise factor expression.

\subsection{Our expressions follow Bangera's noise factor formulas for cascade networks} 
\label{subsec3_2}

Bangera's total noise factor formula for an $n$-stage cascade network is defined as the product of all its stage-wise noise factors, which is a correction to Friis' total noise factor formula \cite{bib15}, given by, 

\begin{equation}
\label{eqn_15}
F_{\text{T}_n}^{\text{Bang}} = \prod_{x=1}^{n}F_x^{\text{Bang}} 
\end{equation}

Comparing equations~\eqref{eqn_11} and \eqref{eqn_15}, the total excess noise factor for a staircase APD is equal to the product of all its stepwise excess noise factors, similar to Bangera's formula for cascade networks. Moreover, Illustrations 1, 2, and 3 in Appendix~\ref{secA3} prove that the total excess noise factor values obtained using our expression (equation~\eqref{eqn_14}) are equal to those estimated using Bangera's formula for cascade networks (equation~\eqref{eqn_15}).

Since the externally added noises are negligible for staircase APDs, Bangera's formula for stage-wise noise factor for cascade networks (with stage-wise internal noises) is given by \cite{bib15}, 

\begin{equation}
\label{eqn_16}
\begin{aligned}
F_x^{\text{Bang}} &= 1+\delta_{\text{int}(x)} \left(\geq 1 \right)   
\end{aligned}
\end{equation}

where $\delta_{\text{int}(x)}$ is the noise component corresponding to the stage-wise internal noises generated due to the irregularities within the $x$-th stage of the cascade network. Comparing equations~\eqref{eqn_13} and \eqref{eqn_16}, the stage-wise (or stepwise) internal noise component in a cascade network (such as staircase APDs) as a function of their stepwise ionization probabilities or the mean and variance of their stepwise gains will be,  

\begin{equation}
\label{eqn_17}
\begin{aligned}
\delta_{\text{int}(x)} = \frac{p_x(1-p_x)}{(1+p_x)^2} = \frac{\text{var}(m_x)}{\langle m_x \rangle ^2}
\end{aligned}
\end{equation}

For statistically independent events of stepwise impact ionization in a staircase APD, the variance of its staircase gain (total stepwise gain) is,

\begin{equation}
\label{eqn_18}
\begin{aligned} 
\text{var}{(m_\text{S})} = \langle m_\text{S}^2 \rangle - \langle m_\text{S} \rangle^2 = \prod_{x=1}^{n} \langle m_x^2 \rangle - \prod_{x=1}^{n} \langle m_x \rangle ^2
\end{aligned}
\end{equation} 

Using equations~\eqref{eqn_15}$-$\eqref{eqn_18}, we can mathematically prove that Bangera's total noise factor expression for cascade networks in terms of the mean and variance of their stage-wise gains is equivalent to the total excess noise factor expression for a staircase APD given by equation~\eqref{eqn_11}, that is, 

\begin{equation}
\label{eqn_19}
\begin{aligned}
F_{\text{T}_n}^{\text{Bang}} = \prod_{x=1}^{n} \left\{ 1+\frac{\text{var}(m_x)}{\langle m_x \rangle^2} \right\} = 1+\frac{\text{var}(m_\text{S})}{\langle m_\text{S} \rangle^2} = F(M_{\text{S}})
\end{aligned}
\end{equation}

The above illustrations and comparisons prove that our excess noise factor expressions match Bangera's formulas. Importantly, the above total (excess) noise factor expressions for cascade networks (such as staircase APDs) are similar to $n$-stage two-port networks connected in cascade where its overall transmission parameters is equal to the product of all its individual (stage-wise) transmission parameters \cite{bib16}.

\section{Conclusion} 
\label{sec4}

We initially formulated generalized expressions for layer-wise and total excess noise factors for multilayer graded-bandgap APDs in terms of their random layer-wise ionization probabilities, applicable for all operating biases. Subsequently, we derived simplified expressions for stepwise and total excess noise factors for multistep staircase APDs in terms of their stepwise ionization probabilities. Remarkably, our newly derived total excess noise factors for a staircase APD can be expressed as the product of its stepwise excess noise factors, similar to Bangera's total noise factor formula for a cascade network \cite{bib15}. We further presented some important expressions for stepwise excess noise factors for staircase APDs as a function of the mean and variance of their stepwise gains. 

In the results and discussion, the curves for total excess noise factor as a function of `$p$' depicted that the curves' peaks remained fixed at $p \approx 0.3$, indicating that the internal noises due to the randomness in the stepwise impact ionization is maximum at $p \approx 0.3$. Additionally, our expression's estimated values agreed well with the measured total excess noise factor data for 3-step staircase APDs reported by March \textit{et al.} (Corrections \& amendments) \cite{bib2} and Dadey \textit{et al.} \cite{bib11}. We further proved that our expressions matched Bangera's noise factor formulas for cascade networks, which are a correction to Friis' formulas, especially for multistage cascade networks with the number of stages $n\geq2$ \cite{bib15}. Collectively, we conclude that our newly derived excess noise factor expressions for staircase APDs are a correction to Capasso's expression and its equivalents. Future work will focus on developing a generalized noise model for cascade networks that can be used in the analysis and simulation of any cascade-amplifying structure in the fields of microelectronics \& photonics \cite{bib2,bib11,bib17,bib18}, two-dimensional optoelectronics \cite{bib19,bib20}, integrated circuits \cite{bib21,bib22,bib23}, chemical biology \cite{bib24,bib25}, artificial intelligence \cite{bib26,bib27,bib28}, and so on.

\section*{Acknowledgments}
A.E.B. would like to thank the Indian Institute of Technology Bombay for their support.

\section*{Appendices}
\appendix

\section{Derivation of excess noise factor expressions for staircase APDs and their proofs} 
\label{secA1}

Consider that an electron at the input of step `$x$' would generate only one free-electron with stepwise ionization probability $p_x$ via impact ionization. Then, $X_x$ is a random variable for carrier (electron) multiplication at step `$x$' with just two possibilities 0 or 1 with probabilities $(1-p_x)$ or $p_x$, respectively. Therefore,

\begin{equation}
\label{Appendix_A_eqn_1}
\begin{aligned}
X_x \sim 
\begin{cases}
0 & (1-p_{x}) \\
& [i.e., \text{No Ionization at step `}x\text{'} \implies X_x=0 \text{ with probability } (1-p_x)]\\
1 & p_{x} \\ 
& [i.e., \text{Ionization at step `}x\text{'} \ni \text{1 e$^-$ generated with probability } p_x]\\
\end{cases}
\end{aligned}
\end{equation}

In this derivation we consider the stepwise impact ionization events as statistically independent. 

\subsection{Solution for a 1-step staircase APD}

\begin{equation}
\label{Appendix_A_eqn_2}
\begin{aligned}
(1+X_1) \sim 
\begin{cases}
1  & (1-p_{1}) \\
2  & p_{1} \\ 
\end{cases}
\end{aligned}
\end{equation}

and 

\begin{equation}
\label{Appendix_A_eqn_3}
\begin{aligned}
(1+X_1)^2 \sim 
\begin{cases}
1  & (1-p_{1}) \\
4  & p_{1} \\ 
\end{cases}
\end{aligned}
\end{equation}

Then, 

\begin{equation}
\label{Appendix_A_eqn_4}
\begin{aligned}
\langle m_\text{S}\rangle ^2 &= \Big\langle (1+X_1) \Big\rangle ^2 \\
&= (1-p_1+2p_1)^2 \\
&= (1+p_1)^2
\end{aligned}
\end{equation}

and 

\begin{equation}
\label{Appendix_A_eqn_5}
\begin{aligned}
\langle m_\text{S}^2\rangle &= \Big\langle (1+X_1)^2 \Big\rangle \\
&= 1-p_1 + 4p_1 \\
&= 1+3p_1 
\end{aligned}
\end{equation}

Thus, the total excess noise factor expression for 1-step staircase APDs is, 

\begin{equation}
\label{Appendix_A_eqn_6}
\begin{aligned}
F(M_{\text{S}},p_1) &= \frac{\langle m_\text{S}^2\rangle}{\langle m_\text{S}\rangle ^2} \\
&= \frac{(1+3p_1)}{(1+p_1)^2} \\ 
&= \frac{\langle m_1^2 \rangle}{\langle m_1 \rangle ^2} \\
&= F_1(M_{\text{S}},p_1)
\end{aligned}
\end{equation} 

Since the above expression for $F(M_{\text{S}},p_1)$ is for a single step APD, this can be considered as an expression for stepwise excess noise factors for a staircase APD. Therefore, the expression for stepwise excess noise factor at the $x$-th step of a staircase APD with stepwise ionization probability $p_x$ can be expressed as, 

\begin{equation}
\label{Appendix_A_eqn_7}
\begin{aligned}
F_x(M_{\text{S}},p_x) = \frac{(1+3p_x)}{(1+p_x)^2} = \frac{\langle m_x^2 \rangle}{\langle m_x \rangle ^2}
\end{aligned}
\end{equation}

But the variance of the stepwise gain is given by,

\begin{equation}
\label{Appendix_A_eqn_8}
\begin{aligned}
\text{var}(m_x) &= \langle m_x^2 \rangle - \langle m_x \rangle ^2 \\
&= (1+3p_x) - (1+p_x)^2 \\ 
&= p_x(1-p_x) 
\end{aligned}
\end{equation} 

Therefore, rearranging equation~\eqref{Appendix_A_eqn_7}, we express the stepwise excess noise factors in terms of mean and variance of their stepwise gains, given by, 

\begin{equation}
\label{Appendix_A_eqn_9}
\begin{aligned}
F_x(M_{\text{S}},p_x) &= 1 + \frac{p_x(1-p_x)}{(1+p_x)^2} \\
&= 1 + \frac{\text{var}(m_x)}{\langle m_x \rangle ^2} \\
&= F_x(M_x)
\end{aligned}
\end{equation}

If the ionization probability $p_1=p$, let the stepwise excess noise factor at the 1-st step of the staircase APD be $F_1(M_{\text{S}},p)=F$. Then, the expression for the total excess noise factor for 1-step staircase APDs with equal stepwise ionization probabilities `$p$' is given by, 

\begin{equation}
\label{Appendix_A_eqn_10}
F(M_{\text{S}},p) = \frac{(1+3p)}{(1+p)^2} = F_1(M_{\text{S}},p) = F 
\end{equation} 

Further, the expression for the stepwise excess noise factor at the $x$-th step of a staircase APD with equal stepwise ionization probabilities `$p$' will be,

\begin{equation}
\label{Appendix_A_eqn_11}
F_x(M_{\text{S}},p)  = \frac{(1+3p)}{(1+p)^2} = F
\end{equation}

\subsection{Solution for a 2-step staircase APD}

\begin{equation}
\label{Appendix_A_eqn_12}
\begin{aligned}
\Bigl((1+X_1)(1+X_2) \Bigr) \sim 
\begin{cases}
1 & (1-p_1)(1-p_2) \\
2 & p_1(1-p_2) \\ 
2 & (1-p_1)p_2 \\
4 & p_1p_2 \\ 
\end{cases}
\end{aligned}
\end{equation}

and 

\begin{equation}
\label{Appendix_A_eqn_13}
\begin{aligned}
\Bigl((1+X_1)^2(1+X_2)^2 \Bigr) \sim 
\begin{cases}
1 & (1-p_1)(1-p_2) \\
4 & p_1(1-p_2) \\ 
4 & (1-p_1)p_2 \\
16 & p_1p_2 \\ 
\end{cases}
\end{aligned}
\end{equation}

Then,

\begin{equation}
\label{Appendix_A_eqn_14}
\begin{aligned}
\langle m_\text{S}\rangle ^2 &= \Big\langle (1+X_1)(1+X_2) \Big\rangle ^2 \\
&= \Bigl[(1-p_1)(1-p_2) + 2p_1(1-p_2) + 2(1-p_1)p_2 + 4p_1p_2 \Bigr]^2 \\
&= (1+p_1)^2(1+p_2)^2 \\ 
&= \prod_{x=1}^{2} (1+p_x)^2 
\end{aligned}
\end{equation}

and 

\begin{equation}
\label{Appendix_A_eqn_15}
\begin{aligned}
\langle m_\text{S}^2\rangle &= \Big\langle (1+X_1)^2(1+X_2)^2 \Big\rangle \\
&= (1-p_1)(1-p_2) + 4p_1(1-p_2) + 4(1-p_1)p_2 + 16p_1p_2 \\
&= 1+3 \{p_1 + p_2 \} + 9 \{ p_1p_2 \} \\ 
&= (1+3p_1)(1+3p_2) \\ 
&= \prod_{x=1}^{2} (1+3p_x) 
\end{aligned}
\end{equation} 

Thus, the total excess noise factor expression for 2-step staircase APDs is, 

\begin{equation}
\label{Appendix_A_eqn_16}
\begin{aligned}
F(M_{\text{S}},p_x) &= \frac{\langle m_\text{S}^2\rangle}{\langle m_\text{S}\rangle ^2} \\ 
&= \prod_{x=1}^{2} \biggl\{ \frac{(1+3p_x)}{(1+p_x)^2} \biggr\} \\ 
&= \prod_{x=1}^{2} F_x(M_{\text{S}},p_x)
\end{aligned}
\end{equation}

\subsection{Solution for a 3-step staircase APD}

\begin{equation}
\label{Appendix_A_eqn_17}
\begin{aligned}
\begin{aligned}
\Bigl((1+X_1) (1+X_2) \\
(1+X_3) \Bigr) 
\end{aligned}
\sim 
\begin{cases}
1 & (1-p_1)(1-p_2)(1-p_3) \\
2 & p_1(1-p_2)(1-p_3) \\ 
2 & (1-p_1)p_2(1-p_3) \\
2 & (1-p_1)(1-p_2)p_3 \\
4 & p_1p_2(1-p_3) \\
4 & p_1(1-p_2)p_3 \\ 
4 & (1-p_1)p_2p_3 \\
8 & p_1p_2p_3 \\
\end{cases}
\end{aligned}
\end{equation}

and 

\begin{equation}
\label{Appendix_A_eqn_18}
\begin{aligned}
\begin{aligned}
\Bigl((1+X_1)^2 (1+X_2)^2 \\
(1+X_3)^2 \Bigr) 
\end{aligned}
\sim 
\begin{cases}
1 & (1-p_1)(1-p_2)(1-p_3) \\
4 & p_1(1-p_2)(1-p_3) \\ 
4 & (1-p_1)p_2(1-p_3) \\
4 & (1-p_1)(1-p_2)p_3 \\
16 & p_1p_2(1-p_3) \\
16 & p_1(1-p_2)p_3 \\ 
16 & (1-p_1)p_2p_3 \\
64 & p_1p_2p_3 \\
\end{cases}
\end{aligned}
\end{equation}

Then, 

\begin{equation}
\label{Appendix_A_eqn_19}
\begin{aligned}
\langle m_\text{S}\rangle ^2 &= \Big\langle (1+X_1)(1+X_2)(1+X_3) \Big\rangle ^2 \\ 
&= \Bigl[(1-p_1)(1-p_2)(1-p_3) + 2p_1(1-p_2)(1-p_3) \\
&~+ 2(1-p_1)p_2(1-p_3) + 2(1-p_1)(1-p_2)p_3 \\
&~+ 4p_1p_2(1-p_3) + 4p_1(1-p_2)p_3 \\
&~+ 4(1-p_1)p_2p_3 + 8p_1p_2p_3 \Bigr]^2 \\
&= (1+p_1)^2(1+p_2)^2(1+p_3)^2 \\ 
&= \prod_{x=1}^{3} (1+p_x)^2 
\end{aligned}
\end{equation}

and 

\begin{equation}
\label{Appendix_A_eqn_20}
\begin{aligned}
\langle m_\text{S}^2\rangle &= \Big\langle (1+X_1)^2(1+X_2)^2(1+X_3)^2 \Big\rangle \\
&= (1-p_1)(1-p_2)(1-p_3) + 4p_1(1-p_2)(1-p_3) \\
&~+ 4(1-p_1)p_2(1-p_3) + 4(1-p_1)(1-p_2)p_3 \\
&~+ 16p_1p_2(1-p_3) + 16p_1(1-p_2)p_3 \\
&~+ 16(1-p_1)p_2p_3 + 64p_1p_2p_3 \\
&= 1+3 \{p_1 + p_2 + p_3 \} + 9 \{ p_1p_2 + p_1p_3 + p_2p_3\} \\
&~+ 27 \{ p_1p_2p_3 \} \\ 
&= (1+3p_1)(1+3p_2)(1+3p_3) \\ 
&= \prod_{x=1}^{3} (1+3p_x) 
\end{aligned}
\end{equation}

Thus, the total excess noise factor expression for 3-step staircase APDs is, 

\begin{equation}
\label{Appendix_A_eqn_21}
\begin{aligned}
F(M_{\text{S}},p_x) &= \frac{\langle m_\text{S}^2\rangle}{\langle m_\text{S}\rangle ^2}  \\
&= \prod_{x=1}^{3} \biggl\{ \frac{(1+3p_x)}{(1+p_x)^2} \biggr\} \\
&= \prod_{x=1}^{3} F_x(M_{\text{S}},p_x)
\end{aligned}
\end{equation}

\subsection{Solution for an \textit{n}-step staircase APD}

\begin{equation}
\label{Appendix_A_eqn_22}
\begin{aligned}
\prod_{x=1}^{n}(1+X_x) \sim 
\begin{cases}
1 & \prod_{x=1}^{n}(1-p_x) \\
2 & p_1\prod_{\forall x;x\neq 1}(1-p_x) \\ 
2 & p_2\prod_{\forall x;x\neq 2}(1-p_x) \\
\vdots & \vdots \\
2 & p_n\prod_{\forall x;x\neq n}(1-p_x) \\
4 & p_1p_2\prod_{\forall x;x\neq 1,2}(1-p_x) \\
4 & p_1p_3\prod_{\forall x;x\neq 1,3}(1-p_x) \\ 
\vdots & \vdots \\
(2)^n & \prod_{x=1}^{n}p_x \\
\end{cases}
\end{aligned}
\end{equation}

and 

\begin{equation}
\label{Appendix_A_eqn_23}
\begin{aligned}
\prod_{x=1}^{n}(1+X_x)^2 \sim 
\begin{cases}
1 & \prod_{x=1}^{n}(1-p_x) \\
4 & p_1\prod_{\forall x;x\neq 1}(1-p_x) \\ 
4 & p_2\prod_{\forall x;x\neq 2}(1-p_x) \\
\vdots & \vdots \\
4 & p_n\prod_{\forall x;x\neq n}(1-p_x) \\
16 & p_1p_2\prod_{\forall x;x\neq 1,2}(1-p_x) \\
16 & p_1p_3\prod_{\forall x;x\neq 1,3}(1-p_x) \\ 
\vdots & \vdots \\
(2^2)^n & \prod_{x=1}^{n}p_x \\
\end{cases}
\end{aligned}
\end{equation}

Then, the expression for mean staircase gain (mean of the total step-gains) for $n$-step staircase APDs is,

\begin{equation}
\label{Appendix_A_eqn_24}
\begin{aligned}
\langle m_\text{S}\rangle &= \Big\langle \prod_{x=1}^{n}(1+X_x) \Big\rangle \\
&= (1+p_1)(1+p_2)(1+p_3) ... (1+p_n) \\
&= \prod_{x=1}^{n}(1+p_x) 
\end{aligned}
\end{equation} 

The square of the mean staircase gain for $n$-step staircase APDs will be, 

\begin{equation}
\label{Appendix_A_eqn_25}
\begin{aligned}
\langle m_\text{S}\rangle ^2 &= \Big\langle \prod_{x=1}^{n}(1+X_x) \Big\rangle ^2 \\
&= (1+p_1)^2(1+p_2)^2(1+p_3)^2 ... (1+p_n)^2 \\
&= \prod_{x=1}^{n}(1+p_x)^2
\end{aligned}
\end{equation}

And the mean square value of staircase gain for $n$-step staircase APDs will be, 

\begin{equation}
\label{Appendix_A_eqn_26}
\begin{aligned}
\langle m_\text{S}^2\rangle &= \bigg\langle \prod_{x=1}^{n}(1+X_x)^2 \bigg\rangle \\
&= 1+ (3)^1 \Biggl\{\sum_{j_1=1}^{n} p_{j_1} \Biggr\} + (3)^2 \Biggl\{\sum_{j_1=1}^{n-1} p_{j_1} \Biggl[ \sum_{j_2=j_1+1}^{n} p_{j_2} \Biggr]\Biggr\} \\
&\hspace{1cm} + (3)^3 \Biggl\{\sum_{j_1=1}^{n-2} p_{j_1} \Biggl[ \sum_{j_2=j_1+1}^{n-1} p_{j_2} \Biggl( \sum_{j_3=j_2+1}^{n} p_{j_3} \Biggr) \Biggr]\Biggr\} \\ 
&\hspace{1cm} + ... + (3)^n \Biggl\{\prod_{j_n=1}^{n} p_{j_n} \Biggr\} \\ 
&= (1+3p_1)(1+3p_2)(1+3p_3) ... (1+3p_n) \\ 
&= \prod_{x=1}^{n} (1+3p_x)
\end{aligned}
\end{equation} 

Therefore, our total excess noise factor expression for $n$-step staircase APDs will be, 

\begin{equation}
\label{Appendix_A_eqn_27}
\begin{aligned}
F(M_{\text{S}},p_x) &= \frac{\langle m_\text{S}^2\rangle}{\langle m_\text{S}\rangle ^2}  \\
&= \prod_{x=1}^{n} \biggl\{ \frac{(1+3p_x)}{(1+p_x)^2} \biggr\} \\
&= \prod_{x=1}^{n} F_x(M_{\text{S}},p_x)
\end{aligned}
\end{equation}

Further, if ionization probabilities at all the steps are equal \textit{i.e.}, $\forall x$; $p_x=p$, then, all the stepwise excess noise factors will also be equal \textit{i.e.}, $\forall x$; $F_x=F$. Therefore, our total excess noise factor expression for $n$-step staircase APDs with equal stepwise ionization probabilities `$p$' can be expressed as, 

\begin{equation}
\label{Appendix_A_eqn_28}
F(M_{\text{S}},p) = \left[ \frac{(1+3p)}{(1+p)^{2}} \right]^n = \Bigl( F_x(M_{\text{S}},p)\Bigr)^n = F^n
\end{equation}

\section{Generalized excess noise factor expressions for multilayer graded-bandgap APDs}
\label{secA2}

Let $X_x$ be a random variable for carrier (electron) multiplication at layer `$x$' that represents the number of free-electrons generated by a single input electron at each layer with variable layer-wise ionization probabilities $p_{xi}$, where `$x$' indicates the layer number and `$i$' indicates the impact-ionized electron count, such that,

\begin{equation}
\label{Appendix_B_eqn_1}
\begin{aligned}
X_x \sim 
\begin{cases}
0 & 1-p_{x1}-p_{x2}-...-p_{xm} \\
1 & p_{x1} \\ 
2 & p_{x2} \\
\vdots & \vdots \\
m & p_{xm} \\
\end{cases}
\end{aligned}
\end{equation}

Then, 

\begin{equation}
\label{Appendix_B_eqn_2}
\begin{aligned}
(1+X_x) \sim 
\begin{cases}
1 & 1-p_{x1}-p_{x2}-...-p_{xm} \\
2 & p_{x1} \\ 
3 & p_{x2} \\
\vdots & \vdots \\
(m+1) & p_{xm} \\
\end{cases}
\end{aligned}
\end{equation}

and 

\begin{equation}
\label{Appendix_B_eqn_3}
\begin{aligned}
(1+X_x)^2 \sim 
\begin{cases}
1 & 1-p_{x1}-p_{x2}-...-p_{xm} \\
4 & p_{x1} \\ 
9 & p_{x2} \\
\vdots & \vdots \\
(m+1)^2 & p_{xm} \\
\end{cases}
\end{aligned}
\end{equation}

We further derive the expressions for multilayer graded-bandgap APDs, similar to the procedures followed in Appendix~\ref{secA1}. Then, the generalized expression for mean of the total layer-gain for $n$-layer graded-bandgap APDs is,

\begin{equation}
\label{Appendix_B_eqn_4}
\begin{aligned}
\langle m_{\text{L}} \rangle &= \bigg\langle \prod_{x=1}^{n} m_x \bigg\rangle \\
&= \bigg\langle \prod_{x=1}^{n} (1+X_x) \bigg\rangle \\
&= \prod_{x=1}^{n} \Bigl( (1-p_{x1}-p_{x2}-p_{x3}-...-p_{xm}) + 2p_{x1} + 3p_{x2} + ... +(m+1)p_{xm} \Bigr) \\
&= \prod_{x=1}^{n} (1+p_{x1}+2p_{x2}+...+mp_{xm}) \\
&= \prod_{x=1}^{n} \biggl( 1+ \sum_{i=1}^{m} \Bigl\{ (i)p_{xi} \Bigr\} \biggr) 
\end{aligned}
\end{equation} 

The square of the mean total layer-gain for $n$-layer graded-bandgap APDs will be, 

\begin{equation}
\label{Appendix_B_eqn_5}
\begin{aligned}
\langle m_{\text{L}} \rangle ^2 &= \bigg\langle \prod_{x=1}^{n} (1+X_x) \bigg\rangle ^2 \\
&= \prod_{x=1}^{n} \Bigl( (1-p_{x1}-p_{x2}-p_{x3}-...-p_{xm}) + 2p_{x1} + 3p_{x2} + ... +(m+1)p_{xm} \Bigr)^2 \\
&= \prod_{x=1}^{n} (1+p_{x1}+2p_{x2}+...+mp_{xm})^2 \\
&= \prod_{x=1}^{n} \biggl( 1+ \sum_{i=1}^{m} \Bigl\{ (i)p_{xi} \Bigr\} \biggr)^2 
\end{aligned}
\end{equation} 

And the mean square value of the total layer-gain for $n$-layer graded-bandgap APDs will be, 

\begin{equation}
\label{Appendix_B_eqn_6}
\begin{aligned}
\langle m_{\text{L}}^2 \rangle &= \bigg\langle \prod_{x=1}^{n} (1+X_x)^2 \bigg\rangle \\
&= \prod_{x=1}^{n} \Bigl( (1-p_{x1}-p_{x2}-p_{x3}-...-p_{xm}) + 4p_{x1} + 9p_{x2} + ... +(m+1)^2p_{xm} \Bigr) \\
&= \prod_{x=1}^{n} (1+3p_{x1}+8p_{x2}+...+[m(m+2)]p_{xm}) \\
&= \prod_{x=1}^{n} \biggl( 1+ \sum_{i=1}^{m} \Bigl\{ [i(i+2)]p_{xi} \Bigr\} \biggr) 
\end{aligned}
\end{equation} 

Therefore, our generalized expression for total excess noise factor for $n$-layer graded-bandgap APDs is, 

\begin{equation}
\label{Appendix_B_eqn_7}
\begin{aligned}
F(M_{\text{L}},p_{xi}) &= \frac{\langle m_\text{L}^2\rangle}{\langle m_\text{L}\rangle ^2}  \\
&= \prod_{x=1}^{n} \Biggl\{ \frac{\Bigl( 1+ \sum_{i=1}^{m} \bigl\{ [i(i+2)]p_{xi} \bigr\} \Bigr)}{\Bigl( 1+ \sum_{i=1}^{m} \bigl\{ (i)p_{xi} \bigr\} \Bigr)^2} \Biggr\} \\ 
&= \prod_{x=1}^{n} F_x(M_{\text{L}},p_{xi}) 
\end{aligned}
\end{equation} 

where $F_x(M_{\text{L}},p_{xi})$ is the stepwise excess noise factor at the $x$-th layer of $n$-layer graded-bandgap APD, and its generalized expression is given by, 

\begin{equation}
\label{Appendix_B_eqn_8}
\begin{aligned}
F_x(M_{\text{L}},p_{xi}) = \frac{\Bigl( 1+ \sum_{i=1}^{m} \bigl\{ [i(i+2)]p_{xi} \bigr\} \Bigr)}{\Bigl( 1+ \sum_{i=1}^{m} \bigl\{ (i)p_{xi} \bigr\} \Bigr)^2} 
\end{aligned}
\end{equation}

\section{Illustrations: Our expressions follow Bangera's noise factor formulas for cascade networks} 
\label{secA3}

Bangera's total noise factor formula for $n$-stage cascade networks, which is a correction to Friis' total noise factor formula \cite{bib15} is given by, 

\begin{equation}
\label{Appendix_C_eqn_1}
F_{\text{T}_n}^{\text{Bang}} = \prod_{x=1}^{n}F_x^{\text{Bang}} 
\end{equation}

For a single-stage network, the equation~\eqref{Appendix_C_eqn_1} reduces to, 

\begin{equation}
\label{Appendix_C_eqn_2}
F_{\text{T}_1}^{\text{Bang}} = F_1^{\text{Bang}} 
\end{equation}

Bangera's formula for stage-wise noise factor at the $x$-th stage of cascade network (with internal noises) 
\cite{bib15} is,

\begin{equation}
\label{Appendix_C_eqn_3}
\begin{aligned}
F_x^{\text{Bang}} &= 1+\frac{\left(N_{\text{int}(x)}+N_{\text{ext}(x)}\right)}{N_i\prod_{j=1}^{x}G_j+\sum_{k=1}^{(x-1)}\left\{\left(N_{\text{int}(k)}+N_{\text{ext}(k)}\right)\prod_{l=k+1}^{x}G_l\right\}} \\ 
&= 1+\frac{N_{\text{int}(x)}}{N_{\text{i}(x)}G_x}+\frac{N_{\text{ext}(x)}}{N_{\text{i}(x)}G_x} \\
&= 1+\delta_{\text{int}(x)}+\frac{N_{\text{ext}(x)}}{N_{i(x)}G_x}
\end{aligned}
\end{equation}

where $N_{\text{int}(x)}$ is the noise power corresponding to the stage-wise internal noises generated due to irregularities within the cascade network's $x$-th stage (or $x$-th step of staircase APDs); $\delta_{\text{int}(x)}$ is a noise component corresponding to the stage-wise internal noises generated due to the irregularities within the $x$-th stage of cascade network (or $x$-th step of staircase APDs); $N_{\text{ext}(x)}$ is the externally added noise power at the output of the $x$-th stage of cascade network; $N_{i(x)}$ is the total noise power at the input of the $x$-th stage of cascade network; and $G_x$ is the power gain at the $x$-th stage of cascade network (or stage-wise power gains).

For staircase APDs, externally added noises are negligible, \textit{i.e.}, $N_{\text{ext}(x)}\approx0$. Then from equation~\eqref{Appendix_C_eqn_3}, we get,  

\begin{equation}
\label{Appendix_C_eqn_4}
\begin{aligned}
F_x^{\text{Bang}} = 1+\delta_{\text{int}(x)} = F_x 
\end{aligned}
\end{equation}

From equation~\eqref{Appendix_A_eqn_27} in Appendix~\ref{secA1}, the total excess noise factor expression for $n$-step staircase APDs is, 

\begin{equation}
\label{Appendix_C_eqn_5}
\begin{aligned}
F(M_{\text{S}},p_x) = \prod_{x=1}^{n} F_x(M_{\text{S}},p_x)
\end{aligned}
\end{equation} 

Comparing equations~\eqref{Appendix_C_eqn_1} and \eqref{Appendix_C_eqn_5}, the total excess noise factor for a staircase APD is equal to the product of all its stepwise excess noise factors, similar to cascade networks.

From equations~\eqref{Appendix_A_eqn_28} in Appendix~\ref{secA1}, the total excess noise factor expression for $n$-step staircase APDs with equal stepwise ionization probabilities `$p$' is,
 
\begin{equation}
\label{Appendix_C_eqn_6}
F(M_{\text{S}},p) = \left[\frac{(1+3p)}{(1+p)^{2}}\right]^n = \Bigl( F_x(M_{\text{S}},p)\Bigr)^n = F^n = F_{\text{T}_n} 
\end{equation}

Here, the mean staircase gain in terms of `$p$' will be $\langle m_\text{S} \rangle=(1+p)^n$ and the mean stepwise gains will be $\langle m_x \rangle =(1+p)$. \\

Bellow are some illustrations that estimate the values of stepwise and total excess noise factors for staircase APDs using equation~\eqref{Appendix_C_eqn_6} and compares these values with those estimated using Bangera's noise factor formulas given by equations~\eqref{Appendix_C_eqn_1}$-$\eqref{Appendix_C_eqn_4}. Considering equal stepwise ionization probabilities as `$p$' at all steps of an $n$-step staircase APD, its stepwise excess noise factors will also be equal at all the steps (\textit{i.e.}, $\forall x$; $F_x=F$). \\ 

\textbf{\underline{Case}:} Let us consider the presence of internal noise, \textit{i.e.}, when the stepwise ionization probabilities `$p$' are less than unity and non-zero. \\

\textbf{Illustration 1:} If $p=0.3$ and $n=1$, then substituting the values of `$p$' and `$n$' in equation~\eqref{Appendix_C_eqn_6}, we get the total excess noise factor for 1-step staircase APD as $F_{\text{T}_1}=1.12426$ (which is greater than unity and agrees with equations~\eqref{Appendix_C_eqn_2}, \eqref{Appendix_C_eqn_3}, and \eqref{Appendix_C_eqn_4} where $F_1^{\text{Bang}}=1+\delta_{int(1)}$ ($F_1^{\text{Bang}} > 1$) when $N_{\text{ext}(1)}=0$). Since this corresponds to a staircase APD with only one-step, the stepwise excess noise factor at step one of this 1-step staircase APD is $F_1=F=F_{\text{T}_1}=1.12426=F_{\text{T}_1}^{\text{Bang}}$. \\ 

\textbf{Illustration 2:} Similarly, if $p=0.3$ and $n=2$, the total excess noise factor for 2-step staircase APD will be $F_{\text{T}_2}=1.26396$. Since we have considered the same values of `$p$' as in the previous illustration, we get $F_2=F_1=F=1.12426$. Substituting these values in Bangera's equation~\eqref{Appendix_C_eqn_1}, we get $F_{\text{T}_2}^{\text{Bang}} = F_1F_2 = 1.26396 = F_{\text{T}_2}$. \\

\textbf{Illustration 3:} Similarly, if $p=0.3$ and $n=3$, the total excess noise factor for 3-step staircase APD will be $F_{\text{T}_3}=1.42102$.  Since we have considered the same values of `$p$' as in the previous illustration, we get $F_3=F_2=F_1=F=1.12426$. Substituting these values in Bangera's equation~\eqref{Appendix_C_eqn_1}, we get $F_{\text{T}_3}^{\text{Bang}} = F_1F_2F_3 = 1.42102 = F_{\text{T}_3}$. \\


\begin{thebibliography}{10}
\expandafter\ifx\csname url\endcsname\relax
  \def\url#1{\burl{#1}}\fi
\expandafter\ifx\csname urlprefix\endcsname\relax\def\urlprefix{URL }\fi
\providecommand{\bibinfo}[2]{#2}
\providecommand{\eprint}[2][]{\url{#2}}
\providecommand{\doi}[1]{\url{https://doi.org/#1}}


\bibitem{bib1}
\bibinfo{author}{David, J.}
\newblock \bibinfo{title}{Photodetectors: {T}he staircase photodiode}.
\newblock \emph{\bibinfo{journal}{Nat. Photon.}}
  \textbf{\bibinfo{volume}{10}}~(6), \bibinfo{pages}{364--366}
  (\bibinfo{year}{2016}).
\newblock \doi{10.1038/nphoton.2016.98} .

\bibitem{bib2}
\bibinfo{author}{March, S.~D.}, \bibinfo{author}{Jones, A.~H.},
  \bibinfo{author}{Campbell, J.~C.} \& \bibinfo{author}{Bank, S.~R.}
\newblock \bibinfo{title}{Multistep staircase avalanche photodiodes with
  extremely low noise and deterministic amplification}.
\newblock \emph{\bibinfo{journal}{Nat. Photon.}}
  \textbf{\bibinfo{volume}{15}}~(6), \bibinfo{pages}{468--474}
  (\bibinfo{year}{2021}).
\newblock \doi{10.1038/s41566-021-00814-x} .

\bibitem{bib3}
\bibinfo{author}{Ripamonti, G.} \emph{et~al.}
\newblock \bibinfo{title}{Realization of a staircase photodiode: {T}owards a
  solid-state photomultiplier}.
\newblock \emph{\bibinfo{journal}{Nucl. Instrum. Methods Phys. Res. A}}
  \textbf{\bibinfo{volume}{288}}~(1), \bibinfo{pages}{99--103}
  (\bibinfo{year}{1990}).
\newblock \doi{10.1016/0168-9002(90)90471-H} .

\bibitem{bib4}
\bibinfo{author}{Teich, M.}, \bibinfo{author}{Matsuo, K.} \&
  \bibinfo{author}{Saleh, B.}
\newblock \bibinfo{title}{Excess noise factors for conventional and
  superlattice avalanche photodiodes and photomultiplier tubes}.
\newblock \emph{\bibinfo{journal}{IEEE J. Quantum Electron.}}
  \textbf{\bibinfo{volume}{22}}~(8), \bibinfo{pages}{1184--1193}
  (\bibinfo{year}{1986}).
\newblock \doi{10.1109/JQE.1986.1073137} .

\bibitem{bib5}
\bibinfo{author}{Capasso, F.}, \bibinfo{author}{Tsang, W.-T.} \&
  \bibinfo{author}{Williams, G.}
\newblock \bibinfo{title}{Staircase solid-state photomultipliers and avalanche
  photodiodes with enhanced ionization rates ratio}.
\newblock \emph{\bibinfo{journal}{IEEE Trans. Electron Devices}}
  \textbf{\bibinfo{volume}{30}}~(4), \bibinfo{pages}{381--390}
  (\bibinfo{year}{1983}).
\newblock \doi{10.1109/T-ED.1983.21132} .

\bibitem{bib6}
\bibinfo{author}{Williams, G.}, \bibinfo{author}{Capasso, F.} \&
  \bibinfo{author}{Tsang, W.}
\newblock \bibinfo{title}{The graded bandgap multilayer avalanche photodiode: {A}
  new low-noise detector}.
\newblock \emph{\bibinfo{journal}{IEEE Electron Device Lett.}}
  \textbf{\bibinfo{volume}{3}}~(3), \bibinfo{pages}{71--73}
  (\bibinfo{year}{1982}).
\newblock \doi{10.1109/EDL.1982.25483} .

\bibitem{bib7}
\bibinfo{author}{Ren, M.} \emph{et~al.}
\newblock \bibinfo{title}{{AlInAsSb}/{GaSb} staircase avalanche photodiode}.
\newblock \emph{\bibinfo{journal}{Appl. Phys. Lett.}}
  \textbf{\bibinfo{volume}{108}}~(8), \bibinfo{pages}{081101}
  (\bibinfo{year}{2016}).
\newblock \doi{10.1063/1.4942370} .

\bibitem{bib8}
\bibinfo{author}{Pilotto, A.} \emph{et~al.}
\newblock \bibinfo{title}{A new expression for the gain-noise relation of
  single-carrier avalanche photodiodes with arbitrary staircase multiplication
  regions}.
\newblock \emph{\bibinfo{journal}{IEEE Trans. Electron Devices}}
  \textbf{\bibinfo{volume}{66}}~(4), \bibinfo{pages}{1810--1814}
  (\bibinfo{year}{2019}).
\newblock \doi{10.1109/TED.2019.2900743} .

\bibitem{bib9}
\bibinfo{author}{van~der Ziel, A.}, \bibinfo{author}{Yu, Y.},
  \bibinfo{author}{Bosman, G.} \& \bibinfo{author}{Van~Vliet, C.}
\newblock \bibinfo{title}{Two simple proofs of {C}apasso's excess noise factor
  ${F}_{N}$ of an ideal ${N}$-stage staircase multiplier}.
\newblock \emph{\bibinfo{journal}{IEEE Trans. Electron Devices}}
  \textbf{\bibinfo{volume}{33}}~(11), \bibinfo{pages}{1816--1817}
  (\bibinfo{year}{1986}).
\newblock \doi{10.1109/T-ED.1986.22746} .

\bibitem{bib10}
\bibinfo{author}{Matsuo, K.}, \bibinfo{author}{Teich, M.} \&
  \bibinfo{author}{Saleh, B.}
\newblock \bibinfo{title}{Noise properties and time response of the staircase
  avalanche photodiode}.
\newblock \emph{\bibinfo{journal}{IEEE Trans. Electron Devices}}
  \textbf{\bibinfo{volume}{32}}~(12), \bibinfo{pages}{2615--2623}
  (\bibinfo{year}{1985}).
\newblock \doi{10.1109/T-ED.1985.22392} .

\bibitem{bib11}
\bibinfo{author}{Dadey, A.~A.}, \bibinfo{author}{Jones, A.~H.},
  \bibinfo{author}{March, S.~D.}, \bibinfo{author}{Bank, S.~R.} \&
  \bibinfo{author}{Campbell, J.~C.}
\newblock \bibinfo{title}{Near-unity excess noise factor of staircase avalanche photodiodes}.
\newblock \emph{\bibinfo{journal}{Optica}}
  \textbf{\bibinfo{volume}{10}}~(10), \bibinfo{pages}{1353--1357} (\bibinfo{year}{2023}).
\newblock \doi{10.1364/OPTICA.496587} .

\bibitem{bib12}
\bibinfo{author}{Friis, H.}
\newblock \bibinfo{title}{Noise figures of radio receivers}.
\newblock \emph{\bibinfo{journal}{Proc. IRE}}
  \textbf{\bibinfo{volume}{32}}~(7), \bibinfo{pages}{419--422}
  (\bibinfo{year}{1944}).
\newblock \doi{10.1109/JRPROC.1944.232049} .

\bibitem{bib13}
\bibinfo{author}{Haus, H.}
\newblock \bibinfo{title}{The noise figure of optical amplifiers}.
\newblock \emph{\bibinfo{journal}{IEEE Photon. Technol. Lett.}}
  \textbf{\bibinfo{volume}{10}}~(11), \bibinfo{pages}{1602--1604}
  (\bibinfo{year}{1998}).
\newblock \doi{10.1109/68.726763} .

\bibitem{bib14}
\bibinfo{author}{Bangera, A.~E.}
\newblock \bibinfo{title}{Mathematical proof of errors in {C}apasso's excess noise factor formula for an $n$-step staircase multiplier}
  (\bibinfo{year}{2025}).
\newblock \bibinfo{note}{{P}reprint at
	\url{https://doi.org/10.48550/arXiv.2506.09649}}. 

\bibitem{bib15}
\bibinfo{author}{Bangera, A.~E.}
\newblock \bibinfo{title}{Corrections to {F}riis noise factor formulas for cascade networks}
  (\bibinfo{year}{2025}).
\newblock \bibinfo{note}{{P}reprint at
	\url{https://doi.org/10.48550/arXiv.2506.09900}}. 



\bibitem{bib16}
\bibinfo{author}{Alexander, C.~K.} \& \bibinfo{author}{Sadiku, M.~N.~O.}
\newblock \bibinfo{title}{Fundamentals of Electric Circuits}.
  \bibinfo{edition}{{7}th edn}.
\newblock	\bibinfo{publisher}{McGraw-Hill Education}, \bibinfo{location}{New York}
  (\bibinfo{year}{2021}).




\bibitem{bib17}
\bibinfo{author}{Terazzi, R.} \emph{et~al.}
\newblock \bibinfo{title}{Bloch gain in quantum cascade lasers}.
\newblock \emph{\bibinfo{journal}{Nat. Phys.}}
  \textbf{\bibinfo{volume}{3}}~(5), \bibinfo{pages}{329--333}
  (\bibinfo{year}{2007}).
\newblock \doi{10.1038/nphys577} .

\bibitem{bib18}
\bibinfo{author}{Nevou, L.} \emph{et~al.}
\newblock \bibinfo{title}{Current quantization in an optically driven electron pump based on self-assembled quantum dots}.
\newblock \emph{\bibinfo{journal}{Nat. Phys.}}
  \textbf{\bibinfo{volume}{7}}~(5), \bibinfo{pages}{423--427}
  (\bibinfo{year}{2011}).
\newblock \doi{10.1038/nphys1918} .



\bibitem{bib19}
\bibinfo{author}{Yang, L.} \emph{et~al.}
\newblock \bibinfo{title}{Efficient photovoltage multiplication in carbon nanotubes}.
\newblock \emph{\bibinfo{journal}{Nat. Photon.}}
  \textbf{\bibinfo{volume}{5}}~(11), \bibinfo{pages}{672--676}
  (\bibinfo{year}{2011}).
\newblock \doi{10.1038/nphoton.2011.250} .

\bibitem{bib20}
\bibinfo{author}{Tielrooij, K.~J.} \emph{et~al.}
\newblock \bibinfo{title}{Photoexcitation cascade and multiple hot-carrier generation in graphene}.
\newblock \emph{\bibinfo{journal}{Nat. Phys.}}
  \textbf{\bibinfo{volume}{9}}~(4), \bibinfo{pages}{248--252}
  (\bibinfo{year}{2013}).
\newblock \doi{10.1038/nphys2564} .



\bibitem{bib21}
\bibinfo{author}{Liu, J}, \bibinfo{author}{Zhang, D.},
  \bibinfo{author}{Wang, M.}, \bibinfo{author}{Huang, L.} \&
  \bibinfo{author}{Zhao, D.}
\newblock \bibinfo{title}{A cascaded linear high-voltage amplifier circuit for dielectric measurement}.
\newblock \emph{\bibinfo{journal}{IEEE Trans. Ind. Electron.}}
  \textbf{\bibinfo{volume}{63}}~(3), \bibinfo{pages}{1834--1841}
  (\bibinfo{year}{2016}).
\newblock \doi{10.1109/TIE.2015.2498129} .

\bibitem{bib22}
\bibinfo{author}{Hackett, L.} \emph{et~al.}
\newblock \bibinfo{title}{Non-reciprocal acoustoelectric microwave amplifiers with net gain and low noise in continuous operation}.
\newblock \emph{\bibinfo{journal}{Nat. Electron.}}
  \textbf{\bibinfo{volume}{6}}~(1), \bibinfo{pages}{76--85}
  (\bibinfo{year}{2023}).
\newblock \doi{10.1038/s41928-022-00908-6} .

\bibitem{bib23}
\bibinfo{author}{Qiu, J.~Y.} \emph{et~al.}
\newblock \bibinfo{title}{Broadband squeezed microwaves and amplification with a {Josephson} travelling-wave parametric amplifier}.
\newblock \emph{\bibinfo{journal}{Nat. Phys.}}
  \textbf{\bibinfo{volume}{19}}~(5), \bibinfo{pages}{706--713}
  (\bibinfo{year}{2023}).
\newblock \doi{10.1038/s41567-022-01929-w} .



\bibitem{bib24}
\bibinfo{author}{Macfarlane, R.~G.}
\newblock \bibinfo{title}{An enzyme cascade in the blood clotting mechanism, and its function as a biochemical amplifier}.
\newblock \emph{\bibinfo{journal}{Nature}}
  \textbf{\bibinfo{volume}{202}}~(4931), \bibinfo{pages}{498--499}
  (\bibinfo{year}{1964}).
\newblock \doi{10.1038/202498a0} .

\bibitem{bib25}
\bibinfo{author}{Wan, X.} \emph{et~al.}
\newblock \bibinfo{title}{Cascaded amplifying circuits enable ultrasensitive cellular sensors for toxic metals}.
\newblock \emph{\bibinfo{journal}{Nat. Chem. Biol.}}
  \textbf{\bibinfo{volume}{15}}~(5), \bibinfo{pages}{540--548}
  (\bibinfo{year}{2019}).
\newblock \doi{10.1038/s41589-019-0244-3} .



\bibitem{bib26}
\bibinfo{author}{Hussain, S.}, \bibinfo{author}{Mokhtar, M.}, \& 
  \bibinfo{author}{Howe, J.~M.}
\newblock \bibinfo{title}{Sensor failure detection, identification, and accommodation using fully connected cascade neural network}.
\newblock \emph{\bibinfo{journal}{IEEE Trans. Ind. Electron.}}
  \textbf{\bibinfo{volume}{62}}~(3), \bibinfo{pages}{1683--1692}
  (\bibinfo{year}{2015}).
\newblock \doi{10.1109/TIE.2014.2361600} .

\bibitem{bib27}
\bibinfo{author}{Sabokrou, M.}, \bibinfo{author}{Fayyaz, M.},
  \bibinfo{author}{Fathy, M.} \& \bibinfo{author}{Klette, R.}
\newblock \bibinfo{title}{Deep-cascade: {C}ascading {3D} deep neural networks for fast anomaly detection and localization in crowded scenes}.
\newblock \emph{\bibinfo{journal}{IEEE Trans. Image Process.}}
  \textbf{\bibinfo{volume}{26}}~(4), \bibinfo{pages}{1992--2004}
  (\bibinfo{year}{2017}).
\newblock \doi{10.1109/TIP.2017.2670780} .

\bibitem{bib28}
\bibinfo{author}{Pan, Z.} \emph{et~al.}
\newblock \bibinfo{title}{{MIEGAN}: {M}obile image enhancement via a multi-module cascade neural network}.
\newblock \emph{\bibinfo{journal}{IEEE Trans. Multimed.}}
  \textbf{\bibinfo{volume}{24}}, \bibinfo{pages}{519--533}
  (\bibinfo{year}{2022}).
\newblock \doi{10.1109/TMM.2021.3054509} .



\end{thebibliography}
\end{document}